\documentclass[final]{svjour3}
\usepackage{graphicx}
\usepackage{rotating}
\usepackage{amssymb}
\usepackage{amsmath}
\usepackage{mathptmx}
\usepackage[numbers]{natbib}
\usepackage{bm}

\makeatletter
\journalname{Journal of Low Temperature Physics}

\def\lesssim{\ \raise.3ex\hbox{$<$}\kern-0.8em\lower.7ex\hbox{$\sim$}\ }
\def\gesim{\ \raise.3ex\hbox{$>$}\kern-0.8em\lower.7ex\hbox{$\sim$}\ }
\bibpunct{}{}{,}{s}{}{,}
\begin{document}
\newcommand{\hdblarrow}{H\makebox[0.9ex][l]{$\downdownarrows$}-}
\title{Momentum distribution of Cooper-pairs and strong-coupling effects in a two-dimensional Fermi gas near the Berezinskii-Kosterlitz-Thouless transition}
\author{M. Matsumoto \and D. Inotani \and Y. Ohashi}
\institute
{\at
Faculty of Science and Technology, Keio University, 3-14-1 Hiyoshi, Kohoku-ku, Yokohama 223-8522, Japan. \\
Tel.: +81-45-566-1454 \\
Fax: +81-45-566-1672 \\
\email{moriom@rk.phys.keio.ac.jp}
}
\par
\date{\today}
\maketitle
\keywords{ultracold Fermi gas, two-dimensional system, BKT phase transition}
\par
\begin{abstract}
We investigate strong-coupling properties of a two-dimensional ultracold Fermi gas in the normal state. Including pairing fluctuations within the framework of a $T$-matrix approximation, we calculate the distribution function $n({\bm Q})$ of Cooper pairs in terms of the center of mass momentum ${\bm Q}$. In the strong-coupling regime, $n({\bm Q}=0)$ is shown to exhibit a remarkable increase with decreasing the temperature in the low temperature region, which agrees well with the recent experiment on a two-dimensional $^6$Li Fermi gas [M. G. Ries, {\it et. al.}, Phys. Rev. Lett. {\bf 114}, 230401 (2015)]. Our result indicates that the observed remarkable increase of the number of Cooper pairs with zero center of mass momentum can be explained without assuming the Berezinskii-Kosterlitz-Thouless (BKT) transition, when one properly includes pairing fluctuations that are enhanced by the low-dimensionality of the system. Since the BKT transition is a crucial topic in two-dimensional Fermi systems, our results would be useful for the study toward the realization of this quasi-long-range order in an ultracold Fermi gas. 
\par
\noindent PACS numbers: 03.75.Hh, 05.30.Fk, 67.85.Lm.
\end{abstract}
\par
\section{Introduction}
\par
The advantage of an ultracold Fermi gas is the high tunability of various physical parameters\cite{Gurarie,Bloch}. Using a tunable pairing interaction associated with a Feshbach resonance, we can now study a Fermi superfluid from the weak-coupling regime to the strong-coupling limit in a systematic manner\cite{Giorgini}. In addition, the system dimensionality can be also tunable by using an optical lattice potential\cite{Bloch}. Indeed, a two-dimensional Fermi gas has recently been realized by using this technique\cite{Sommer,Feld,Frohlich,Murthy,Ries,Murthy2}. Because of these experimental developments, strong-coupling properties of a two-dimensional Fermi gas has become an interesting and realistic research topic in cold Fermi gas physics\cite{Tempere,Iskin,Klimin,Pietila,Watanabe,Bauer,Matsumoto,Marsiglio,Levinsen}.
\par
In contrast to a three-dimensional Fermi gas, the superfluid long-range order is prohibited in the two-dimensional case, because it is completely destroyed by low dimensional superfluid fluctuations\cite{Mermin,Hohenberg}. However, a two-dimensional Fermi gas is known to be able to still exhibit superfluid properties, when the Berezinskii-Kosterlitz-Thouless (BKT) phase transition occurs\cite{Berezinskii,Kosterlitz}.
\par
Recently, the observation of this quasi-long-range order was reported in a two-dimensional $^6$Li Fermi gas\cite{Ries}. While the BKT transition is theoretically explained on the viewpoint of vortex-antivortex pair annihilation, this experiment determines the BKT phase transition temperature $T_{\rm BKT}^{\rm exp}$ as the temperature below which the number $N_{\rm C}({\bm Q}=0)$ of Cooper pairs with zero center of mass momentum (${\bm Q}=0$) remarkably increases\cite{Ries}. Since the vortex-antivortex pair annihilation is not observed in this experiment\cite{Ries}, it is a crucial issue to check whether or not $T_{\rm BKT}^{\rm exp}$ determined from the temperature dependence of $N_{\rm C}({\bm Q}=0)$ can unambiguously be identified as the BKT transition temperature predicted theoretically.
\par
In this paper, to examine this, we investigate a two-dimensional Fermi gas near $T_{\rm BKT}^{\rm exp}$. We discuss how to evaluate $N_{\rm C}({\bm Q}=0)$ in a strong-coupling $T$-matrix approximation (TMA). Although this strong-coupling theory cannot describe the BKT transition\cite{Schmitt,Tokumitu}, we show that the observed remarkable increase of this quantity below a certain temperature (which is experimentally identified as the BKT transition temperature\cite{Ries}) can be explained without assuming the BKT phase transition. We also present an alternative explanation for this phenomenon on the viewpoint of strong pairing fluctuations that are enhanced by the low-dimensionality of the system. Throughout this paper, we take $\hbar =k_{\rm B}=1$ and the two-dimensional system area is taken to be unity, for simplicity.
\par
\section{Formulation}
\par
We consider a two-dimensional uniform Fermi gas consisting of two atomic hyperfine states, described by the BCS Hamiltonian,
\begin{align}
 H=\sum_{\bm{p},\sigma}\xi_{\bm{p}}c^{\dagger}_{\bm{p},\sigma}c_{\bm{p},\sigma}
  -U\sum_{\bm{p},\bm{p}'\bm{q}}c^{\dagger}_{\bm{p}+\bm{q}/2,\uparrow}c^{\dagger}_{-\bm{p}+\bm{q}/2,\downarrow}
  c_{-\bm{p}'+\bm{q}/2,\downarrow}c_{\bm{p}'+\bm{q}/2,\uparrow}. \label{eq2.1}
\end{align}
Here, $c^{\dagger}_{\bm{p},\sigma}$ is a creation operator of a Fermi atom with  pseudospin $\sigma =\uparrow, \downarrow$ and two-dimensional momentum $\bm{p}=(p_{x},p_{y})$. $\xi_{\bm p}=p^2/(2m)-\mu$ is the kinetic energy, measured from the Fermi chemical potential $\mu$, where $m$ is an atomic mass. The pairing interaction $-U$ ($<0$) is assumed to be tunable by adjusting the threshold energy of a Feshbach resonance. As usual, we measure the interaction strength in terms of the two-dimensional $s$-wave scattering length $a_{2{\rm D}}$, which is related to $U$ as\cite{Morgan} $1/U=(m/2\pi)\ln{(k_{\rm F}a_{2{\rm D}})}+\sum_{p\geq k_{\rm F}}m/p^2$ (where $k_{\rm F}=\sqrt{2\pi N}$ is the Fermi momentum, with $N$ being the total number of Fermi atoms).
\par
\begin{figure}[t]
\begin{center}
\includegraphics[width=1\textwidth]{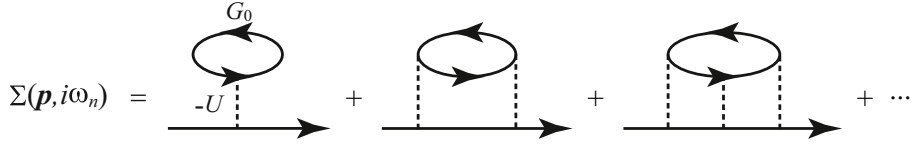}
\caption{
Self-energy $\Sigma({\bm p},i\omega_n)$ in the $T$-matrix approximation (TMA). The solid line represents the non-interacting Green's function $G_0$, and the dashed line denotes the pairing interaction $-U$.}
\label{fig1}       
\end{center}
\end{figure}

Many-body corrections to Fermi single-particle excitations can be conveniently described by the self-energy $\Sigma(\bm{p},i\omega_{n})$ in the single-particle thermal Green's function,
\begin{equation}
G(\bm{p},i\omega_{n})=
{1 \over i\omega_n-\xi_{\bm p}-\Sigma(\bm{p},i\omega_{n})},
\label{eq.1}
\end{equation}
where $\omega_n$ is the fermion Matsubara frequency. In the $T$-matrix approximation, the self-energy $\Sigma(\bm{p},i\omega_{n})$ is diagrammatically described as Fig. \ref{fig1}, which gives\cite{Perali,Tsuchiya},
\begin{equation}
\Sigma({\bm p},i\omega_n)
=T\sum_{\bm{q},i\nu_{n}}
\Gamma(\bm{q},i\nu_{n})G_{0}(\bm{q}-\bm{p},i\nu_{n}-i\omega_{n}).
\label{eq.2.2}
\end{equation}
Here, $\nu_{n}$ is the boson Matsubara frequency, and $G_0^{-1}({\bm p},i\omega_n)=i\omega_n-\xi_{\bm p}$ is the single-particle Green's function in a free Fermi gas. The particle-particle scattering matrix $\Gamma(\bm{q},i\nu_{n})$ in TMA has the form\cite{Perali,Tsuchiya},
\begin{equation}
\Gamma(\bm{q},i\nu_{n})=-\frac{U}{1-U\Pi(\bm{q},i\nu_{n})}, 
\label{eq.2.3}
\end{equation}
where
\begin{equation}
\Pi(\bm{q},i\nu_{n})=T\sum_{\bm{p},i\omega_{n}}G_{0}\left(\bm{p}+\frac{\bm{q}}{2},i\nu_{n}+i\omega_{n}\right)G_{0}\left(-\bm{p}+\frac{\bm{q}}{2},-i\omega_{n}\right)
\label{eq.2.4a}
\end{equation}
is the lowest-order pair-correlation function, describing fluctuations in the Cooper channel. 
\par
\par
The equation for the total number $N$ of Fermi atoms in TMA is given by, 
\begin{equation}
N=2T\sum_{\bm{p},i\omega_{n}}G(\bm{p},i\omega_{n}). 
\label{eq.2.4}
\end{equation}
This number equation may be divided into the sum of the free-fermion contribution,
\begin{equation}
N_0=2T\sum_{\bm{p},i\omega_{n}}G_0(\bm{p},i\omega_{n}),
\label{eq.2.5}
\end{equation}
and the fluctuation correction $\delta N$ described by the TMA self-energy $\Sigma({\bm p},i\omega_n)$ in Eq. (\ref{eq.2.2}),
\begin{equation}
\delta N=2T\sum_{\bm{p},i\omega_{n}}
\left[G(\bm{p},i\omega_{n})-G_{0}(\bm{p},i\omega_{n})\right].
\label{eq.2.5a}
\end{equation}
Pairing fluctuations in the BCS-BEC crossover region are physically understood as the repetition of the formation and dissociation of preformed Cooper pairs, that eventually become tightly bound molecular bosons in the strong-coupling BEC limit. Keeping this in mind, and writing Eq. (\ref{eq.2.5a}) as $\delta N=2\sum_{\bm Q}n({\bm Q})$, one may regard 
\begin{equation}
n({\bm Q})=T\sum_{i\nu_{n}}
\Gamma({\bm Q},i\nu_{n})T\sum_{\bm{p}, i\omega_{n}}
G_{0}(\bm{Q}-\bm{p},i\nu_{n}-i\omega_{n})G_{0}(\bm{p},i\omega_{n})
G(\bm{p},i\omega_{n})
\label{eq.2.6}
\end{equation}
as the number of preformed Cooper pairs with the center of mass momentum ${\bm Q}$. Indeed, in the strong-coupling limit (where $\mu\to -\infty$), $n({\bm Q})$ is reduced to the ordinary momentum distribution in an ideal Bose gas,
\begin{equation}
n({\bm Q})=
n_{\rm  B}\left(\frac{Q^2}{2M}-\mu_{\rm B}\right),
\label{eq.2.7}
\end{equation}
where $n_{\rm B}(\omega)$ is the Bose distribution function, $M=2m$ is a molecular mass, and 
\begin{equation}
\mu_{\rm B}=2\mu \ln{\left(\frac{ 2|\mu| }{ E_{\rm bind } } \right)}~~(<0)
\label{eq.2.8}
\end{equation}
is interpreted as the Bose chemical potential, with $E_{\rm bind}=1/(ma^2_{2{\rm D}}$) being the binding energy of a two-body bound state. Although such a molecular picture gradually becomes worse as one approaches the weak-coupling BCS regime, $n({\bm Q})$ in Eq. (\ref{eq.2.6}) is still a useful quantity to grasp the bosonic character of the system in the BCS-BEC crossover region. In this paper, thus, we identify this momentum distribution function $n({\bm Q})$ with the observed number $N_{\rm C}({\bm Q})$ of Cooper-pair bosons with the center of mass momentum ${\bm Q}$ in a two-dimensional $^6$Li Fermi gas\cite{Ries,Riesmaterial}. 
\par
We briefly note that the two-dimensional $T$-matrix approximation we are using in this paper does not give the BCS-type superfluid phase transition\cite{Mermin,Hohenberg}, and also cannot describe the BKT phase transition\cite{Watanabe,Schmitt,Tokumitu}. Thus, the comparison of our TMA result on $n({\bm Q})$ with the recent experiment\cite{Ries,Riesmaterial} provides a useful information about whether or not the observed anomalous increase of $N_{\rm C}({\bm Q=0})$ is really a clear signature of the BKT transition. 
\par
In this paper, to compare our TMA results with the recent experiment on a two-dimensional $^6$Li Fermi gas\cite{Ries,Riesmaterial}, we take $\ln{(k_{\rm F}a_{2{\rm D}})}= -0.59$. At this interaction strength, we first calculate the chemical potential $\mu(T)$ from the number equation (\ref{eq.2.4}), and then evaluate $n({\bm Q})$ in Eq. (\ref{eq.2.6}) at various temperatures. We briefly note that Ref.\cite{Ries} reports the BKT phase transition temperature $T_{\rm BKT}^{\rm exp}=0.129T_{\rm F}$ at this interaction strength, where $T_{\rm F}$ is the Fermi temperature.
\par
\begin{figure}[t]
\begin{center}
  \includegraphics[width=0.5\textwidth]{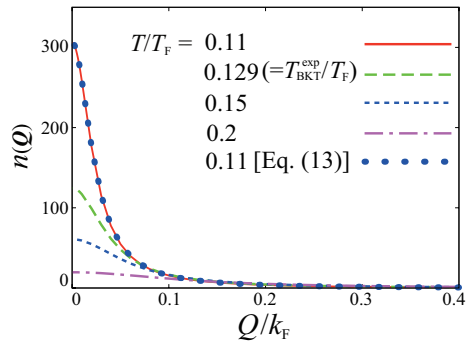}
\caption{(Color online) (a) Calculated momentum distribution function $n({\bm Q})$ of Cooper pairs with respect to the center of mass momentum ${\bm Q}$. We take $\ln{(k_{\rm F}a_{2{\rm D}})}=-0.59$. The dotted line shows the approximate result in Eq. (\ref{eq.3.2}).}
\label{fig2}       
\end{center}
\end{figure}
\par
\section{Momentum distribution function of Cooper pairs}
\par
\begin{figure}[t]
\begin{center}
\includegraphics[width=0.5\textwidth]{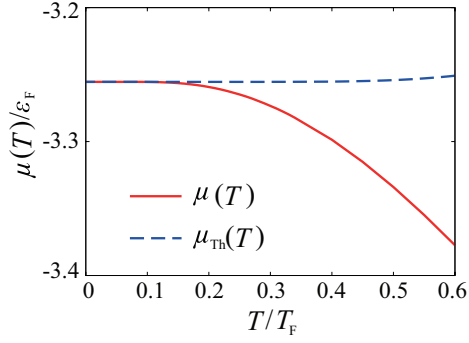}
\caption{(Color online) Calculated Fermi chemical potential $\mu$ as a function of temperature, when $\ln{(k_{\rm F}a_{2{\rm D}})}=-0.59$. $\mu_{\rm Th}$ is the chemical potential which satisfies the Thouless criterion\cite{Thouless}. Although $\mu$ is very close to $\mu_{\rm Th}$ when $T/T_{\rm F}\lesssim 0.15$, the former is always slightly lower than the latter.}
\label{fig3}    
\end{center}
\end{figure}

Figure \ref{fig2} shows the momentum distribution function $n({\bm Q})$ in the low temperature region where the BKT phase transition was experimentally reported ($T_{\rm BKT}^{\rm exp}/T_{\rm F}=0.129$). In this figure, $n({\bm Q})$ in the low momentum region is found to be remarkably enhanced with decreasing the temperature. To see to what extent this behavior reflects bosonic character of the system, it is convenient to approximately evaluate the TMA self-energy $\Sigma({\bm p},i\omega_n)$ in Eq. (\ref{eq.2.2}) by employing the so-called static approximation as\cite{Tsuchiya,Chen},
\begin{equation}
\Sigma ({\bm p},i\omega_n)
\simeq
\left[T\sum_{\bm{q},i\nu_{n}}\Gamma(\bm{q},i\nu_{n})\right]
G_{0}(-\bm{p},-i\omega_{n})
\equiv
-\Delta_{\rm PG}^2G_{0}(-\bm{p},-i\omega_{n}).
\label{eq.3.1}
\end{equation}
Here, $\Delta_{\rm PG}$ is the so-called pseudogap parameter in the literature\cite{Chen}. This approximation assumes that pairing fluctuations described by $\Gamma(\bm{q},i\nu_{n})$ are enhanced in the low-momentum and low-energy region. In addition, as shown in Fig. \ref{fig3}, the Fermi chemical potential $\mu$ is negative and $|\mu|/\varepsilon_{\rm F}>3$ when $\ln{(k_{\rm F}a_{2{\rm D}})}=-0.59$ (where $\varepsilon_{\rm F}$ is the Fermi energy), indicating that the system at this interaction strength is already in the strong-coupling regime. Including this, one can approximate $n({\bm Q})$ in Eq. (\ref{eq.2.6}) to
\begin{equation}
n({\bm Q})
=\frac{-2\mu (\mu+\sqrt{\mu^2+\Delta^2_{\rm PG}})}{\Delta^2_{\rm PG} }
n_{\rm B}\left(\frac{q^2}{2M}-\mu_{\rm B}\right)
={1 \over \displaystyle 1+{2\varepsilon_{\rm F} \over E_{\rm bind}}}
n_{\rm B}\left(\frac{q^2}{2M}-\mu_{\rm B}\right),
\label{eq.3.2}
\end{equation}
where we have used the strong-coupling expression for the pseudogap parameter, $\Delta_{\rm PG}\simeq 2\sqrt{\varepsilon_{\rm F}[\varepsilon_{\rm F}+E_{\rm bind}/2]}$, as well as $\mu\simeq-E_{\rm bind}/2$, in obtaining the last expression (where the two-body binding energy $E_{\rm bind}$ is given below Eq. (\ref{eq.2.8})). As shown in Fig. \ref{fig2}, Eq. (\ref{eq.3.2}) well describes the TMA momentum distribution function $n({\bm Q})$ when $T/T_{\rm F}=0.11$. This clearly indicates that the behavior of $n({\bm Q})$ shown in Fig. \ref{fig2} strongly reflects the bosonic character of this system\cite{note}, originating from the formation of preformed Cooper pairs.
\par
\par
\par
\begin{figure}[t]
\begin{center}
\includegraphics[width=0.5\textwidth]{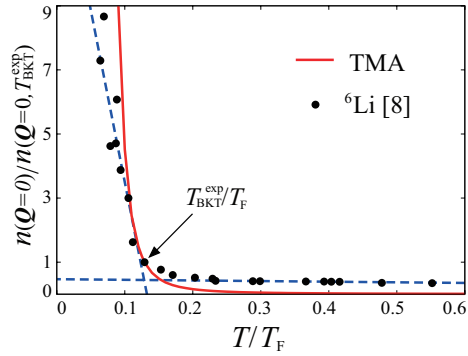}
\caption{(Color online) Calculated momentum distribution function $n({\bm Q})$ at ${\bm Q}=0$ when $\ln{(k_{\rm F}a_{2{\rm D}})}=-0.59$. We also show the recent experiment on the number of Cooper pairs with zero center of mass momentum in a two-dimensional $^6$Li Fermi gas\cite{Ries}. The two dashed lines are experimental fitting lines. Reference \cite{Ries} determines the BKT phase transition temperature $T_{\rm BKT}^{\rm exp}=0.129T_{\rm F}$ as the temperature at which these two lines crosses each other. Each of theoretical and experimental result is normalized by the value at $T_{\rm BKT}^{\rm exp}$. 
}
\label{fig4}       
\end{center}
\end{figure}

Figure \ref{fig4} shows $n({\bm Q}=0)$ when $\ln{(k_{\rm F}a_{2{\rm D}})}=-0.59$. In this figure, we find that the remarkable increase of this quantity around ${\bm Q}=0$ seen in Fig. \ref{fig2} starts to occur when $T\simeq T_{\rm BKT}^{\rm exp}=0.129T_{\rm F}$. We also find that this temperature dependence agrees well with the recent experiment on a two-dimensional $^6$Li Fermi gas\cite{Ries}. As mentioned previously, since our TMA gives no superfluid phase transition around $T_{\rm BKT}^{\rm exp}=0.129T_{\rm F}$, this agreement indicates that the observed remarkable increase of the number of Cooper pairs with zero center of mass momentum does not necessarily mean that the system is in the BKT phase. Further experimental studies would be necessary to confirm that the BKT phase is really realized below $T_{\rm BKT}^{\rm exp}=0.129T_{\rm F}$ at this interaction strength.
\par
\par
\par
The enhancement of $n({\bm Q}=0)$ in TMA is due to strong pairing fluctuations enhanced by the two-dimensionality of the system. To explain this, we expand the denominator of the particle-particle scattering matrix $\Gamma({\bm q},i\nu_n=0)$ in Eq. (\ref{eq.2.3}) around ${\bm q}=0$ as
\begin{equation}
\Gamma({\bm q},i\nu_n=0)\simeq
-{U \over [1-U\Pi({\bm q}=0,i\nu_n=0)]+\alpha q^2},
\label{eq.3.3}
\end{equation}
where $\alpha$ is a positive constant. When the so-called Thouless criterion\cite{Thouless}, $1-U\Pi(0,0)=0$, is satisfied (which is known as the condition for the BCS-type superfluid phase transition in the three-dimensional case), the self-energy in Eq. (\ref{eq.3.1}) diverges in the two-dimensional case, reflecting the logarithmic divergence of the momentum integration of Eq. (\ref{eq.3.3}) [$\sim \int qdq (1/q^2$)]\cite{Schmitt,Tokumitu}. As a result, the chemical potential $\mu$ never reaches the value ($\equiv\mu_{\rm Th}$) that satisfies the Thouless criterion. (See Fig. \ref{fig3}.) Because of this, TMA does not give the superfluid phase transition in a two-dimensional Fermi gas. However, in the low temperature region where $\mu\simeq\mu_{\rm Th}$ in Fig. \ref{fig3}, one expects that $1-U\Pi(0,0)\simeq 0$. In this case, since the particle-particle scattering matrix in Eq. (\ref{eq.2.3}) also has the meaning of an interaction between Fermi atoms, this effective pairing interaction is found to remarkably be enhanced by many-body effects in the low-energy and low-momentum region, which positively contributes to the formation of Cooper pairs. As a result, $n({\bm Q}=0)$ in TMA anomalously increases in this temperature region, as shown in Fig. \ref{fig4}.
\par
Before ending this section, we briefly note that Ref.\cite{Ries} reports that $T_{\rm BKT}^{\rm exp}$ is higher for a weaker pairing interaction, which is opposite to the theoretical prediction that the BKT phase transition temperature decreases with decreasing the interaction strength\cite{Tempere,Matsumoto}. In this regard, the temperature dependence of $n({\bm Q}=0)$ in TMA exhibits the same tendency as the experimental result\cite{Ries} (although we do not show the result here), which also implies the importance of pairing fluctuations in considering this physical quantity. 
\par
\section{Summary}
\par
To summarize, we have investigated a two-dimensional ultracold Fermi gas in the BCS-BEC crossover region. Including pairing fluctuations within a $T$-matrix approximation (TMA), we calculated the momentum distribution function $n({\bm Q})$ of Cooper pairs in the normal state. We showed that the anomalous enhancement of this quantity at ${\bm Q}=0$ that have recently been observed in a two-dimensional $^6$Li Fermi gas\cite{Ries} may be quantitatively explained without assuming the BKT phase transition. That is, even in the normal state, strong pairing fluctuations in the BCS-BEC crossover region that are enhanced by the two dimensionality of the system also lead to the enhancement of $n({\bm Q}=0)$ around the temperature which was experimentally identified as the BKT phase transition temperature $T_{\rm BKT}^{\rm exp}$. This indicates that further experimental studies are needed to confirm that the system is really in the BKT phase below $T_{\rm BKT}^{\rm exp}$. Since the BKT transition is one of the most crucial topics in the field of two-dimensional Fermi superfluid, our results would contribute to the study of this quasi-long-range order in cold Fermi gas physics.
\par
\par
\begin{acknowledgements}
We thank M. G. Ries and P. A. Murthy for useful comments, as well as
 sending their experimental data. We also thank R. Hanai, H. Tajima,
 T. Yamaguchi, and P. van Wyk for discussions. M. M. was supported by KLL PhD Program
Research Grant, as well as Graduate School Doctoral Student Aid Program from Keio
University.This work was supported by the KiPAS project in Keio university. Y.O was supported by Grant-in-Aid for Scientific Research from MEXT and JSPS in Japan (No.25400418, No.15H00840).
\end{acknowledgements}
\par
\par


\begin{thebibliography}{99}
\bibitem{note} The prefactor of the Bose distribution function in Eq. (\ref{eq.3.2}) equals 0.76 in the case of Fig. \ref{fig2}. This prefactor is reduced to unity in the BEC limit where $E_{\rm bind}=-\infty$.
\end{thebibliography}
\end{document}